\documentstyle[12pt,amsmath]{article}
\newcommand{\ch}{\mbox{ch}}
\newcommand{\sh}{\mbox{sh}}
\newcommand{\th}{\mbox{th}}
\begin{document}
\begin{center}
{\Large\bf
The maximum\strut{} entropy tecniques and the statistical
description of systems\strut}
\vskip 5mm
B.Z. Belashev${^{1)}}$ and M.K. Suleymanov${^{2,3)}}$
\vskip 5mm

{\small ${^{1)}}$ Institute of Geology, Karelian Research Centre, RAS, Petrozavodsk, Russia\\
${^{2)}}$ LHE JINR, Dubna, Moscow region, Russia\\
${^{3)}}$ NPI ASCR, Rez near Prague, Czech Republic}
\end{center}
\vskip 5mm
\centerline{\bf Abstract}

The maximum entropy technique (MENT) is used to determine the distribution
functions of physical values. MENT naturally combines required maximum
entropy, the properties of a system and connection conditions in the form 
of restrictions imposed on the system. It can, therefore, be employed to 
statistically describe closed and open systems. Examples in which MENT is 
used to describe equilibrium and non-equilibrium states, as well as steady 
states that are far from being in thermodynamic equilibrium, are discussed.
\medskip

Statistical characteristics are widely used in high-energy physics to analyse
experimental data~\cite{[1]}. A simple analytical technique is employed by 
estimating the entropy of nuclear reaction products – a parameter which 
is sensitive to order-disorder transition, and provides evidence for 
the formation of ordered systems during a reaction~\cite{[2]}.

Entropy methods for simulation of processes are being increasingly 
applied to various processes~\cite{[3]}. An example of such an approach is the 
study of dynamic symmetry and internal chaos in models of interacting 
bozons~\cite{[4],[5]} conducted using the maximum entropy technique
(MENT)~\cite{[6]}.

MENT provides an efficient tool for solving inverse problems~\cite{[7]}. Its 
application to statistical problems is promising. It is desirable, 
however, to test MEM for models that have already been proven to be 
reliable for statistical descriptions. In the present study, the 
applicability of MENT was tested by describing the equilibrium and 
non-equilibrium state of systems. 

The goal of a statistical description of a system is to determine the 
distribution functions of physical values represented by an integral 
image in the results of observation. If distribution functions are 
assessed only on the basis of observation data on a system, then, 
from a mathematical point of view, an incorrect inverse problem arises 
in which the requirements of the existence, uniqueness and stability 
of a solution are not met. The condition of maximum entropy makes it 
possible to meet this requirement automatically. Only one solution of 
the problem is selected out of all. This solution has a maximum entropy, 
is most probable and, therefore, most stable. The resultant solution 
is obviously positive because the MENT estimate is in exponential form 
and agrees with one of the main properties of the distribution function 
--- non-negativity. Other properties of the distribution function, e.g. 
normalization in MENT, can be preset as additional restrictions.

The theory of MENT is based on the second law of thermodynamics which 
describes the thermodynamic equilibrium of a closed system by maximum 
entropy. Attempts have been made to generalize this principle by arguing 
that entropy in internal dissipative processes is maximum~\cite{[8]}. The 
requirement of maximum entropy, the property of a system and condition 
of connection in the form of restrictions placed on the system are 
naturally combined in MENT so that it can be used to statistically 
describe the equilibrium and non-equilibrium states of open and 
closed systems.

The mathematical structure of an entropy functional is related to the 
type of the quantum statistics of a physical data carrier and the 
extent of filling of quantum degrees of freedom~\cite{[7]}. The entropy functional
\begin{equation}
H=-\sum\limits_{\xi=1}^N{f(\xi)\ln f(\xi)},
\end{equation}
where $f(\xi)$ is the distribution function estimate and $\xi$ is an
independent variable, is used for fermions and bozons with a low probability
of filling of energy levels, such as photons of incoherent electromagnetic
radiation, hadrons and $\gamma$-quanta. When the filling of energy levels 
is highly probable, a different form of a functional, like the one used 
in radioastronomy or spectral analysis~\cite{[9]}, is selected:
\begin{equation}
H=-\sum\limits_{\xi=1}^N{\ln f(\xi)}
\end{equation}

We will use the most common form of an  entropy functional (1).

\medskip
{\bf 1. Boltzman's distribution}
\medskip

Let us use MENT to determine the distribution of $N$-particles on the 
$K$-levels of energy $E_i$, which has a maximum entropy, total energy $E$
and the number of particles $N$ being retained:
\begin{align}
 &\sum\limits_{i=1}^K {n_i\cdot E_i}=E\nonumber\\[-6pt]
&\\[-6pt]
 &\sum\limits_{i=1}^K {n_i=N}\nonumber 
\end{align}
Let us formulate the problem of estimating the probability
$f_i=\frac{{n_i}}{N}$
of finding a particle on level $E_i$ as a variation problem on a conditional
extremum with the Lagrange factors $\lambda$ and $\mu$ in the form:
\begin{equation}
-\sum\limits_{i=1}^K{f_i\ln f_i}+\lambda\left(NE-\sum_{i=1}^Kf_iE_i\right)
+\mu\left(1-\sum\limits_{i=1}^K{f_i}\right)\to\max.
\end{equation}
Let us equate the first variation of this relation (4) on $f_i$ to zero.
We will then have a relation:
\begin{equation}
-\ln f_i-1-\mu-\lambda E_i=0
\end{equation}
which gives an expression for probability $f_i$:
\begin{equation}
f_i=\frac{1}{Z}\exp(-\lambda E_i).
\end{equation}
This distribution function is identical to the Boltzman distribution
function if factor $\lambda$ is understood as inverse temperature
$\lambda=1/kT$, and factor $\mu=\ln(Z/e)$.

The requirement of maximum entropy in a closed system with a constant 
number of particles and constant energy results in Boltzman distribution.

\medskip
{\bf 2. Non-equilibrium distribution}
\medskip

Let us use the MENT algorithm to describe the state of a two-level system 
(levels 1 and 2) which has magnetic moment $M$ and consists of $N$-particles 
with magnetic moment $m$. Let us consider that a magnetic field is absent,
the system is not in equilibrium and the magnetic moments of the particles 
located on levels 1 and 2 have opposite directions. 

Let $f(1)$ and $f(2)$ are the probabilities of finding a particle on 
levels 1 and 2, respectively. Equations for the complete probability 
and polarization $P$ of the system will be regarded as connection 
conditions:
\begin{equation}
f(1)+f(2)=1
\end{equation}
\begin{equation}
f(1)-f(2)=M/Nm=P.
\end{equation}
To estimate $f(1)$ and $f(2)$, we will use a conventional MENT scheme 
with a functional:
\begin{equation}
-f(1)\ln f(1)-f(2)\ln f(2)+\mu(1-f(1)-f(2))+\lambda(P+f(2)-f(1))\to\max.
\end{equation}

The formulas
\begin{align}
&  - \ln f(1) - 1 - \mu  - \lambda  = 0 \nonumber\\[-6pt]
&\\[-6pt]
&  - \ln f(2) - 1 - \mu  + \lambda  = 0\nonumber
\end{align}
obtained from condition (9) give expressions for probabilities:
\begin{align}
& f(1) = \exp ( - 1 - \mu  - \lambda ) \nonumber\\[-6pt]
&\\[-6pt]
& f(2) = \exp ( - 1 - \mu  + \lambda ).\nonumber
\end{align}
The substitution of these expressions in conditions (7) and (8)
determines a relationship between Lagrange factors:
\begin{align}
& \exp ( - 1 - \mu )2\ch(\lambda ) = 1 \nonumber\\[-6pt] 
&\\[-6pt]
& \exp ( - 1 - \mu )2\sh(\lambda ) = P.\nonumber
\end{align}
Dividing the first equation by the second one, we will have expression:
\begin{equation}
P=\th(\lambda)
\end{equation}
for the polarization $P$ of the system. Comparing it with an expression 
for the polarization of the equilibrium state of a two-level system with 
magnetic moment $M$ in magnetic field $H$~\cite{[10]}:
\begin{equation}
P=\th(mH/kT)
\end{equation}
we conclude that taking $\lambda=mH/kT$, we thus characterize a 
non-equilibrium state as quasi-equilibrium with magnetic field $H$ 
which agrees with given polarization $P$ or magnetic moment $M$:
\begin{equation}
H(P)=(kT/2m)\ln[(1 + P)/(1 - P)].
\end{equation}

The result obtained implies that if there is a magnetic field which is 
switched off instantly, the thermodynamically equilibrium state of a 
two-level system becomes non-equilibrium and cannot change immediately. 
Therefore, at moments of time that are small in comparison with the 
relaxation time of the system this state is described as quasi-equilibrium, 
provided there is a magnetic field which decreases gently in magnitude.

\medskip
{\bf 3. Steady states of open systems}
\medskip

The steady states of open systems that are far from equilibrium are 
observed in many processes. Systems with autocatalytic chemical reactions 
were among the first to show characteristic types of behaviour of such 
systems, such as double stability, oscillation and wave generation 
interpreted by the mechanism of autocatalysis~\cite{[8]}. An example of such a 
reaction is Belousov-Zhabotinsky's reaction with cerium sulphate
$Ce_2(SO_4)_3$,
malonic acid $CH_2(COOH)_2$, potassium bromate $KbrO_3$ and ferroine 
that are vigorously mixed and supplied to reaction volume from which 
reaction products are simultaneously removed. Cerium, which catalyzes 
the reaction, contributes to the production of two molecules from 
one molecule of an intermediate product of $HbrO_2$:
\begin{equation}
HBrO_2+BrO_3^-+3H^++2Ce^{3+}\to 2HBrO_2+2Ce^{4+}+H_2O.
\end{equation}
In this case, the catalyst is not spent, but its valency changes. 
Ferroine adds a red colour to the solution, if there is an excess of 
trivalent cerium ions, and a blue colour if there is an excess of 
quadrivalent cerium ions. If the time of staying of the reagents in 
the reaction zone is too short to adjust direct and inverse reaction 
rates, then such phenomena as bistability and chemical hours (periodic 
alternation of bistable states with an excess of tri- or quadrivalent 
cerium ions) occur in the system. The colour of the reaction solution 
varies with a period of several minutes. If the reagents are not 
mixed, waves are generated and travel. 

To exemplify the steady states of open systems, let us discuss a 
simple model of autocatalysis~\cite{[8]} represented by two conjugate chemical 
reactions:
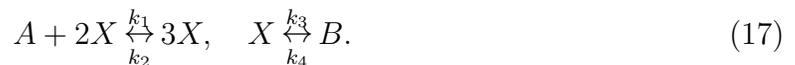
\begin{equation}
A+2X\mathop\leftrightarrow\limits_{k_2 }^{k_1 }3X,\quad
X\mathop\leftrightarrow\limits_{k_4 }^{k_3 }B.
\end{equation}
The concentration of intermediate product $X$ is accepted as the only 
state variable. Substances $A$ and $B$ are either continuously supplied 
to the system or are removed from it, thereby providing the constancy 
of their concentrations $a$ and $b$. Substance $A$ is a catalyst. It 
increases the amount of intermediate product $X$ which, at the same 
time, is transformed to substance $B$.

At equilibrium, the rates of direct and inverse reactions are equal:
\begin{align}
 k_1 ax^2  &= k_2 x^3  \nonumber\\[-6pt]
&\\[-6pt]
 k_3 x &= k_4 b. \nonumber
\end{align}
This relation determines in the only manner the concentration of 
intermediate product $x$ and the reagent concentration ratio $a/b$:
\begin{align}
& x = \frac{{k_4 b}}{{k_3 }} = \frac{{k_1 a}}{{k_2 }} \nonumber\\[-6pt]
&\\[-6pt]
& \frac{b}{a} = \frac{{k_1 k_3 }}{{k_2 k_4 }}. \nonumber 
\end{align}
When a steady state is far from equilibrium, it is characterized by the 
balancing of the total effects of direct and inverse reactions in the 
form of a cubic equation:
\begin{equation}
- k_2 x^3  + k_1 ax^2  - k_3 x + k_4 b = 0.
\end{equation}
Using the conventional MENT scheme to describe such a steady state with 
connection condition (20)
\begin{equation}
 - x\ln x + \lambda ( - k_2 x^3  + k_1 ax^2  - k_3 x + k_4 ) \to \max,
\end{equation}
we come to the equation:
\begin{equation}
\ln x = \lambda ( - 3k_2 x^2  + 2k_1 ax - k_3 )
\end{equation}
which can easily be solved graphically.

It is much more difficult now to trace a relationship between Lagrange 
factor $\lambda$ and concentration $x$ because of a lack of an analytical 
expression. However, the characteristics of the process described can 
be determined from relation (22). This equation can have either one or 
two solutions. In the latter case, bistability exists in an open 
chemical system and the phenomenon of chemical hours can occur. In 
this case, the values of concentrations x corresponding to these 
states are found to be quantitatively dependent and obviously positive 
and the states are most stable in accordance with MENT.

A relationship between Lagrange factor $\lambda$, concentration $x$
and other parameters of the problem can easily be shown for the case 
when equation (22) has one solution. The graphs of the logarithm in 
the left part and those of a parabola in the right part of the equation 
touch each other and, consequently, have a common tangent. Let us 
transform the equality of the derivatives of these functions
\begin{equation}
\frac{1}{x} = \lambda ( - 6k_2 x + 2k_1 a)
\end{equation}
into a quadratic equation:
\begin{equation}
6k_2 \lambda x^2  - 2k_1 a\lambda x + 1 = 0.
\end{equation}
Because the solution is unique, the discriminant of this quadratic 
equation must be equal to zero:
\begin{equation}
4k_1^2 a^2 \lambda ^2  - 24k_2 \lambda  = 0.
\end{equation}
Lagrange factor $\lambda$ can thus be expressed by the relation:
\begin{equation}
\lambda=\frac{{6k_2}}{{k_1^2a^2}},
\end{equation}
and concentration $x$ , which passes to a bistable mode, by the relation:
\begin{equation}
x=\frac{{k_1a}}{{6k_2}}.
\end{equation}
The
value. It has been shown experimentally that transition to a bistable 
mode occurs before thermodynamic equilibrium sets in. 

MENT is a convenient technique for simulation of systems because it 
allows to use not only empirical but also a priori information. One 
can estimate the sensitivity of the distribution function to one or 
another condition. To model complex statistical systems on a computer, 
special MENT algorithms~\cite{[11],[12]} are used.

\medskip\noindent
Conclusions
\medskip

1. MENT is an up-to-date and promising technique, which can efficiently be 
used to solve statistical problems. The MENT estimate of the distribution 
function agrees with the density properties of probability. 
The non-negativity property is satisfied automatically because the MENT 
solution has an exponential form, and normalization can be considered 
one of additional conditions observed when applying the technique.

2. Generalizing the second law of thermodynamics for open systems, 
one can use MENT for their statistical description.  

3. When describing equilibrium and non-equilibrium states on the basis 
of conventional statistical models, MENT gives known results and new 
results when used to describe the steady states of open systems with 
autocatalytic reactions.

4. A simple scheme, observation data and a priori information used as 
restrictions make MENT a convenient tool for describing and modelling 
various processes, including those in high-energy physics.

\end{document}